\documentclass[conference]{IEEEtran}
\IEEEoverridecommandlockouts
\usepackage{cite}
\usepackage{amsmath,amssymb,amsfonts}
\usepackage{algorithmic}
\usepackage{graphicx}
\usepackage{textcomp}
\usepackage{pgfplots}
\pgfplotsset{compat=1.17} % or any other compatible version
\usepackage{xcolor}
\def\BibTeX{{\rm B\kern-.05em{\sc i\kern-.025em b}\kern-.08em
    T\kern-.1667em\lower.7ex\hbox{E}\kern-.125emX}}
\usepackage{tikz}
\usetikzlibrary{shapes.geometric, arrows}

\tikzstyle{block} = [rectangle, draw, fill=blue!20, 
    text width=10em, text centered, rounded corners, minimum height=2em, font=\footnotesize]
\tikzstyle{wideblock} = [rectangle, draw, fill=red!20, 
    text width=22em, text centered, rounded corners, minimum height=2em, font=\footnotesize]
\tikzstyle{arrow} = [thick,->,>=stealth, shorten <=1pt, shorten >=1pt]
\usepackage{url}
\def\BibTeX{{\rm B\kern-.05em{\sc i\kern-.025em b}\kern-.08em
    T\kern-.1667em\lower.7ex\hbox{E}\kern-.125emX}}
\begin{document}

\title{Convolutional neural network for early detection of lameness and irregularity in horses using an IMU sensor\\
{\footnotesize }
\thanks{Innovation project supported by Innosuisse}
}

\author{\IEEEauthorblockN{1\textsuperscript{st} Benoît Savoini*}
\IEEEauthorblockA{\textit{Information Science Institute, GSEM/CUI} \\
\textit{University of Geneva}\\
Carouge, Switzerland \\
benoit.savoini@hesge.ch}
~\\
\and
\IEEEauthorblockN{2\textsuperscript{nd} Jonathan Bertolaccini}
\IEEEauthorblockA{\textit{ Information Science Institute, GSEM/CUI} \\
\textit{University of Geneva}\\
Carouge, Switzerland \\
jonathan.bertolaccini@unige.ch}
~\\
\and
\IEEEauthorblockN{3\textsuperscript{rd} Stéphane Montavon}
\IEEEauthorblockA{\textit{Veterinary Department} \\
\textit{Swiss Armed Forces}\\
Berne, Switzerland\\
smontavon@bluewin.ch}

\and
\IEEEauthorblockN{4\textsuperscript{th} Michel Deriaz}
\IEEEauthorblockA{\textit{Haute Ecole de Gestion Genève} \\
\textit{HES-SO}\\
Carouge, Switzerland \\
michel.deriaz@hesge.ch}
}

\maketitle

\begin{abstract}
Lameness and gait irregularities are significant concerns in equine health management, affecting performance, welfare, and economic value. Traditional observational methods rely on subjective expert assessments, which can lead to inconsistencies in detecting subtle or early-stage lameness. While AI-based approaches have emerged, many require multiple sensors, force plates, or video systems, making them costly and impractical for field deployment.
In this applied research study, we present a stride-level classification system that utilizes a single inertial measurement unit (IMU) and a one-dimensional convolutional neural network (1D CNN) to objectively differentiate between sound and lame horses, with a primary focus on the trot gait. The proposed system was tested under real-world conditions, achieving a 90\% session-level accuracy with no false positives, demonstrating its robustness for practical applications. By employing a single, non-intrusive, and readily available sensor, our approach significantly reduces the complexity and cost of hardware requirements while maintaining high classification performance.
These results highlight the potential of our CNN-based method as a field-tested, scalable solution for automated lameness detection. By enabling early diagnosis, this system offers a valuable tool for preventing minor gait irregularities from developing into severe conditions, ultimately contributing to improved equine welfare and performance in veterinary and equestrian practice.

\end{abstract}


\section{Introduction}
Lameness and gait irregularities in horses are common and important problems in equine health management, affecting performance, welfare and economic value. They are characterized by abnormal gait resulting from pain, functional or structural dysfunction of the musculoskeletal system \cite{b19}. Whether ridden or harnessed, the horse may exhibit gait abnormalities classified as irregularity or lameness according to a widely used lameness scale developed by the AAEP (American Association of Equine Practitioners) \cite{b20}. This scale from 0 to 5 shows irregularity from 0 to 1 and lameness from 1.5 to 5. According to Dyson \textit{et al.} \cite{b8} lameness and/or irregularity are some of the most common reasons for horses to be presented for veterinary evaluation, impacting both leisure and performance horses. Traditionally, the detection and diagnosis of lameness rely heavily on the subjective judgment of veterinarians and experienced handlers, which can vary significantly and may not always be accurate \cite{b17}. This variability highlights the need for objective, reliable, and automated methods to identify lameness early and accurately. 
Advances in sensor technology and data analysis techniques offer promising solutions to this problem. Wearable sensors, such as accelerometers, gyroscopes, and inertial measurement units (IMUs), have been increasingly utilized to collect detailed motion data from horses \cite{b21}. These sensors provide a continuous and minimally invasive means to monitor gait patterns, capturing high-resolution data on the movement of different parts of the horse’s body. Such data are crucial for detecting subtle gait abnormalities that may indicate lameness. In equine studies, several approaches have been developed to harness the power of sensor data. For example, Pasquiet \textit{et al.} \cite{b1} investigated the use of inertial sensors to assess horse locomotion modifications. Similarly, Haladjian \textit{et al.} \cite{b6} used accelerometers to evaluate lameness in dairy cattle, highlighting the advantages of using objective measures over subjective evaluations. However, the challenge lies in effectively analyzing this high-dimensional sensor data to identify subtle deviations from normal gait patterns. Traditional statistical methods may not be sufficient to handle the complexity and volume of the data. Machine learning, particularly deep learning, has shown great potential in processing complex data and uncovering patterns that are not immediately apparent. Convolutional neural networks, a powerful class of deep learning architectures, excel at classification tasks by learning robust features from input data. By capturing hierarchical representations of both normal and abnormal examples, CNNs can effectively differentiate between these classes. This approach has found success in a variety of domains, including machine fault detection \cite{b24} and medical diagnostics \cite{b25}. In the realm of animal health, similar approaches have been explored to detect lameness. For example, Yigit \textit{et al.} \cite{b11} employed accelerometer data for pose estimation as a method for horse lameness detection, demonstrating the effectiveness of machine learning algorithms in processing sensor data to identify lameness with high accuracy. More broadly, existing horse lameness detection techniques have explored a range of approaches, including pose estimation-based methods \cite{b9,b2,b10}, as well as systems relying on multiple sensors \cite{b18} or custom-built hardware \cite{b14}. While these methods have demonstrated effectiveness in identifying gait abnormalities, they often rely on specialized equipment, complex experimental setups, or controlled environments, limiting their feasibility for routine equine health monitoring in real-world conditions. In addition, machine learning approaches were employed to classify lameness in dairy cattle based on automated gait and body movement analysis \cite{b15,b12,b13,b16,b3,b7,b22}. These studies highlight the potential of sensor-based and machine learning methodologies in identifying lameness between different species. Schlageter-Tello \textit{et al.} \cite{b23} also emphasized the importance of automated lameness detection systems to improve animal welfare and management efficiency.

In this study, we propose a novel field-tested system to detect lameness in horses using a supervised learning framework that classifies multivariate time series data collected from a single, readily available sensor attached to the horse. Our method employs a convolutional neural network architecture, specifically designed to learn discriminative temporal features from gait data. The model leverages labeled examples of both sound and lame horses, enabling the CNN to capture meaningful differences between normal and abnormal gait patterns. Unlike previous methods that often rely on multiple sensors, complex setups or custom equipment, this streamlined approach capitalizes on the accessibility of a single, easily available sensor in the market. As a result, our system offers a practical and effective solution for routine lameness detection, supporting timely intervention and improved outcomes for horses.
The structure of the paper is as follows. We begin with a detailed description of our methodology, covering the Alogo sensor used in this study, data collection processes, pre-processing steps, and the architecture of the proposed model. This is followed by the presentation of experimental results that demonstrate the effectiveness of our approach. We then discuss the findings and their implications. Finally, the paper concludes with suggestions for future research directions in this area.

\section{Materials and methods}

\subsection{Sensor}
The Alogo Move Pro sensor is a state-of-the-art wearable device designed to provide comprehensive and precise data on equine movement. Engineered to be lightweight and unobtrusive, it measures approximately 11.5 cm (4.52 in) in length, 2.6 cm (1.02 in) in height, and 5.8 cm (2.28 in) in width, and weighs only 127 g (4.47 oz)—comparable to or even lighter than a typical smartphone. This compact design allows the sensor to be easily attached to the horse’s girth, enabling continuous monitoring without hindering the animal’s natural movement. Equipped with advanced accelerometers and gyroscopes, the Move Pro captures high-resolution, three-dimensional data on the horse’s gait and overall biomechanics, further underscoring its non-intrusive, high-precision capabilities.
The sensor operates by recording the intricate details of the horse's movements, including speed, acceleration, and rotational angles across all three axes. This data is invaluable for identifying subtle abnormalities in gait that may indicate the onset of lameness or other musculoskeletal issues. The device’s robust design ensures reliable performance in various environmental conditions, making it suitable for use in both indoor arenas and outdoor fields.

\begin{figure}[h]
\centering
\includegraphics[scale=0.17]{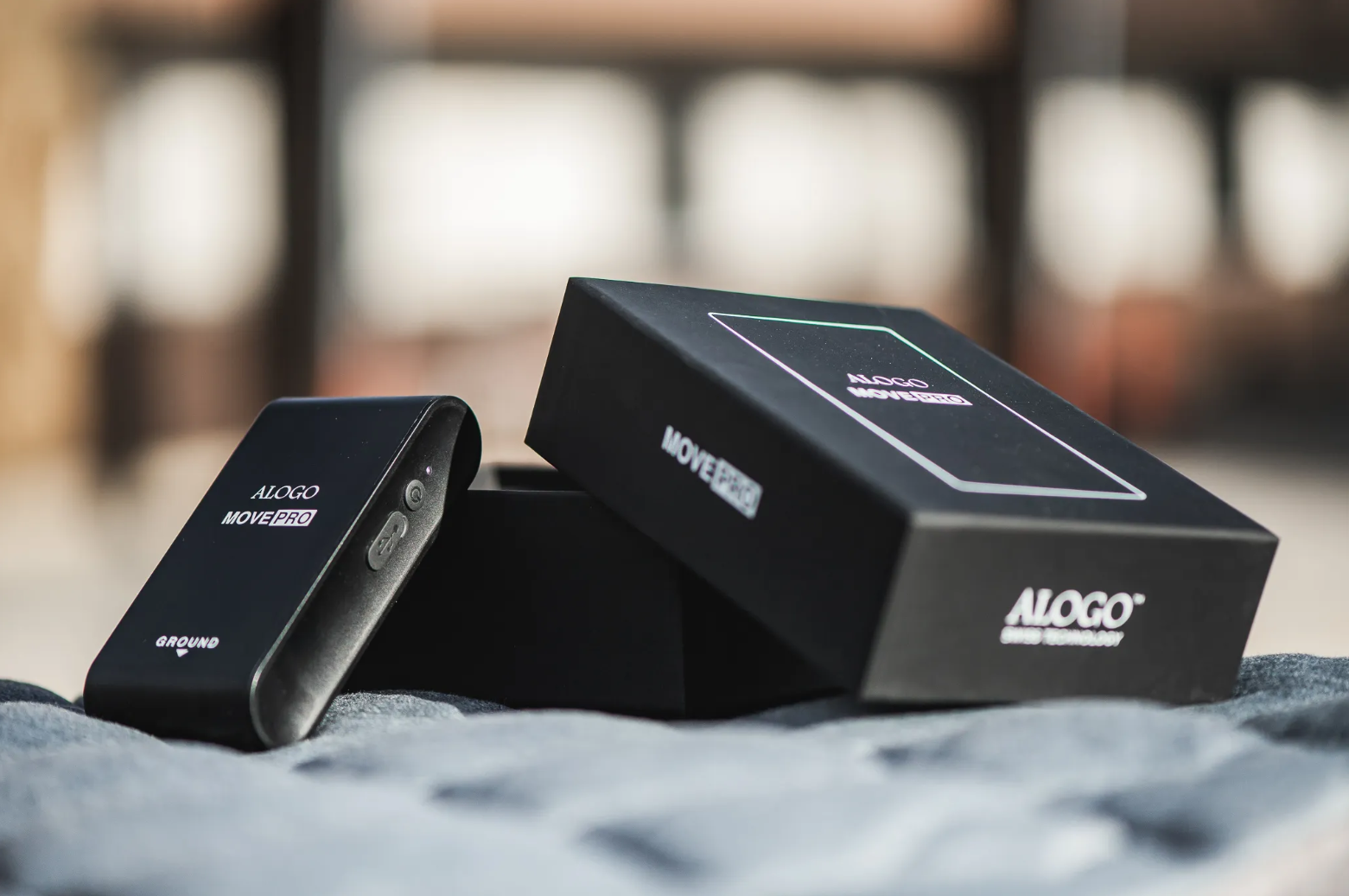}
\caption{Picture of the Alogo Move Pro sensor.}
\end{figure}

\begin{figure}[h]
\centering
\includegraphics[scale=0.17]{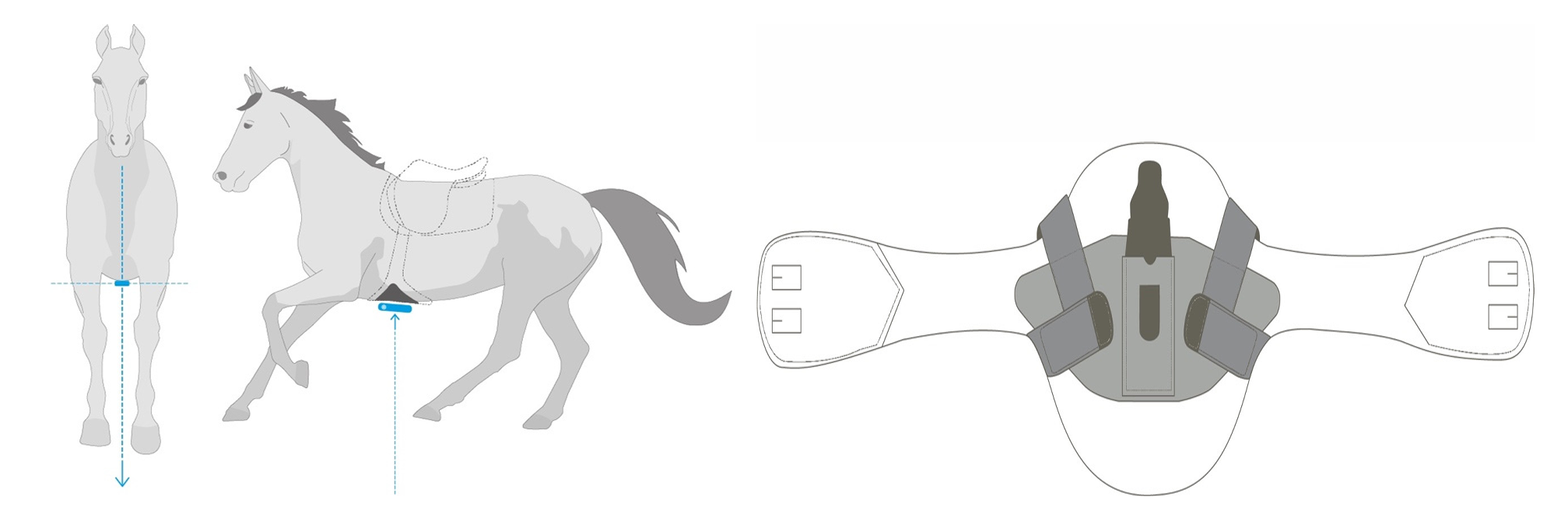}
\caption{Two drawings show the positioning and mounting of the sensor}
\end{figure}

One of the standout features of the Alogo Move Pro is its ability to synchronize with accompanying software to provide real-time analysis and feedback. The collected data is transmitted wirelessly to a user-friendly interface where it is processed and displayed in an easily interpretable format. This enables trainers, veterinarians, and horse owners to make informed decisions based on objective, quantitative data rather than solely relying on subjective visual assessments.

\subsection{Data}

The dataset used in this study comprises data collected with the Alogo Move Pro from 42 horses for a total of 184 sessions collected. We opted for a stride-level classification approach, assigning each stride in a session the same label as the session itself. We acknowledge this is a considerable assumption and do not expect particularly high accuracy at the individual stride scale. However, our overarching focus remains on session-level classification, where examining smaller, more manageable data segments helps the model capture the key characteristics needed for accurate session-wide decisions. Due to the trot’s distinct two-beat pattern, we selected it as the primary gait for lameness detection, setting it apart from other gaits such as the canter, gallop, and walk, which exhibit more complex footfall sequences. This straightforward, symmetrical cadence makes the trot an ideal starting point for our analysis. We detected sessions with low trot stride numbers and put them apart on the assumption that 40 trot strides would be mandatory for the study to ensure reliable classification, thus we only considered 161 sessions for our analysis, as the remaining 23 sessions lacked a sufficient number of trot strides. This decision ensures that the dataset used is robust and focused on sessions with adequate trot data for meaningful analysis. Also, strides containing multiple gaits were excluded from the dataset, ensuring that only complete trot strides were retained for analysis. 
Each horse was equipped with a single sensor placed on the horse's girth, chosen for its focal location to capture comprehensive movement data reflecting the horse's overall gait dynamics. The sensor recorded speed, acceleration, and rotation across the x, y, and z axes at a frequency of 100 Hz, providing a high-resolution, multidimensional view of the horse’s movements. A post-processing step was performed by Alogo, yielding additional variables that describes the horse's movement. Data collection was conducted under controlled conditions to ensure consistency. We built a multivariate time series dataset where each session serves as an input, representing continuous data collected from the sensor. For each session, we considered a total of 7 variables that capture different aspects of the horse's gait dynamics, providing a comprehensive view of its movement patterns. Each session is associated with a binary output, indicating whether it corresponds to a sound or lame horse. This dataset structure allows for effective analysis of gait patterns and supports accurate session-level classification.
After training the model to classify individual strides (sound or lame), we then return to the session level by computing an anomaly score. This score is defined as the ratio of lame-classified trot strides to the total number of trot strides in the session. Consequently, the model’s stride-level predictions are aggregated to produce a more holistic assessment of each session’s overall condition. Rather than using generic overlapping frames, we employed stride segmentation with a maximum stride length of 100 timestamps (1 second at 100 Hz). Through plotting the distribution of trot stride lengths in Fig.~3, and by questioning veterinarians, we established that one second sufficiently covers the majority of real-world stride durations. After segmentation, the stride segments were concatenated and used for both training and evaluation.
To address the issue of class imbalance in our dataset shown in Table. 1, we implemented a class weighting strategy during model training. 

\vspace{0.5cm} % Adjust the vertical spacing as needed

\begin{figure}[h]
\centering
\includegraphics[scale=0.33]{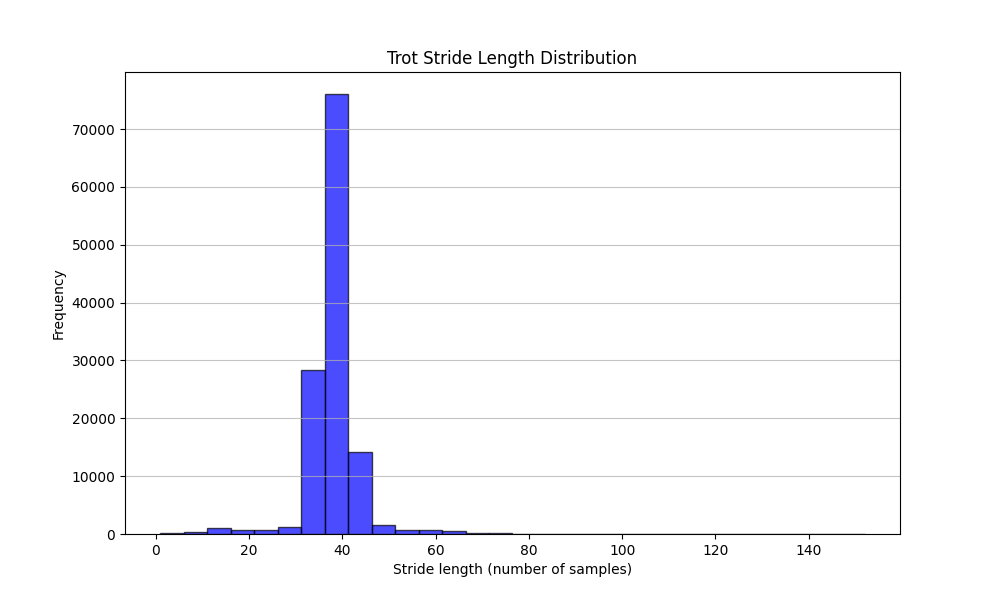}
\caption{Trot stride length distribution of our dataset}
\end{figure}

\begin{table}[h]
\centering
\caption{Number of Sound and Lame Strides per Dataset Split}
\begin{tabular}{lccc}
\hline
\textbf{Dataset} & \textbf{Sound Strides} & \textbf{Lame Strides} & \textbf{Total} \\
\hline
\textbf{Training}   & 48,579 & 23,017 & 71,596 \\
\textbf{Validation} & 14,810 &  14,630 & 29,440 \\
\textbf{Testing}    & 17,746 &  8,237 & 25,983 \\
\hline
\end{tabular}
\label{tab:stride_distribution}
\end{table}

The function calculates weights for each class using the class\textunderscore weight utility from the scikit-learn library. These weights are designed to balance the contribution of each class during training, ensuring that the model does not become biased toward the majority class.
The function takes the training labels as input and computes the weights based on the proportion of stride samples in each class. 
By incorporating these weights, the training process is guided to treat both classes more equally, regardless of their imbalance in the dataset. This approach helps improve the model’s sensitivity to the minority class, enhancing its ability to detect lameness even in underrepresented scenarios. 
Although we considered segmenting our datasets at the horse level, we ultimately chose a session-level division due to numerous confounding factors—such as variability in the horse's condition or environment across sessions—that could alter performance. Indeed, the same horse might yield different results across different sessions, making session-level segmentation a more reliable choice for our analysis. The final train, validation, and test distributions, as well as the specific number of horses and total recorded hours, will be detailed in later sections.

\subsection{Model}

In exploring suitable deep learning methods for analyzing horse gait data, we considered several architectures but ultimately settled on a straightforward 1D CNN. We initially evaluated the use of autoencoders, which reconstruct “normal” data to detect anomalies. However, lameness often manifests over an entire session rather than within short, isolated sequences, making a direct classification approach more effective than reconstruction-based methods. We also investigated LSTM models, which are adept at capturing long-term temporal dependencies but found them prone to overfitting in our dataset.

\vspace{0.5cm} % Adjust the vertical spacing as needed

\begin{center}
\begin{tikzpicture}[node distance=1.4cm, >=stealth, auto]

    % Define a 'block' style for the nodes
    \tikzstyle{block} = [rectangle, draw, 
        text width=12em, text centered, rounded corners, minimum height=2em]

    % Define an 'arrow' style for the edges
    \tikzstyle{arrow} = [->, thick]

    % Nodes
    \node (input) [block, fill=yellow!20] {Input Tensor  (Stride-Level Padded Segments)};
    \node (mask) [block, below of=input] {Masking};    
    \node (conv1) [block, below of=mask] {Conv1D + BatchNorm + ReLU + MaxPool + Dropout};
    \node (conv2) [block, below of=conv1] {Conv1D + BatchNorm + ReLU + MaxPool + Dropout};
    \node (gap) [block, below of=conv2] {Global Average \\ Pooling 1D};
    \node (dense1) [block, below of=gap] {Dense};
    \node (dense) [block, below of=dense1] {Dense \\ (Sigmoid)};

    \node (output) [block, below of=dense, fill=orange!20] {Output \\ (Lameness Class)};

    % Edges
    \draw [arrow] (input) -- (mask);
    \draw [arrow] (mask) -- (conv1);    
    \draw [arrow] (conv1) -- (conv2);
    \draw [arrow] (conv2) -- (gap);
    \draw [arrow] (gap) -- (dense1);
    \draw [arrow] (dense1) -- (dense);    
    \draw [arrow] (dense) -- (output);

\end{tikzpicture}
\vspace{-0.2cm} % Adjust the vertical spacing as needed
\end{center}

\begin{figure}[h]
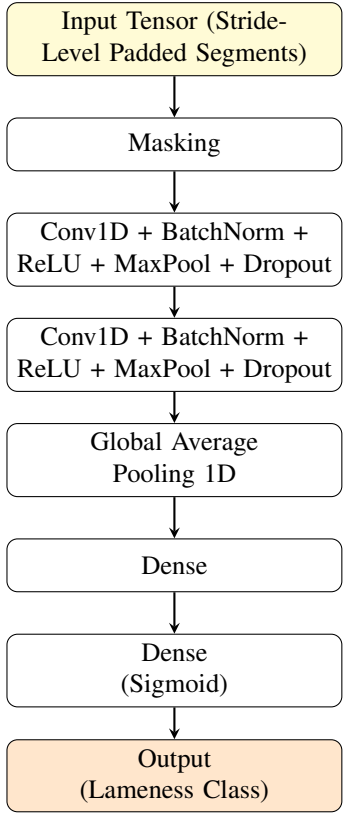

    \centering
    \caption{Overview of the 1D CNN model architecture for lameness detection.}
\end{figure}

As illustrated in Fig.~4, our streamlined 1D CNN processes stride-level segments with fewer parameters, thereby reducing the risk of overfitting while still effectively learning discriminative features for lameness detection. Additionally, strides were padded and masking was employed to facilitate better handling of input data by the model.

\subsection{Parameters}

The model training was conducted using the Adam optimizer. The initial learning rate was set to $10^{-3}$. To refine the learning process, a learning rate scheduler was implemented which reduces the learning rate by a factor of $0.5$ if the validation loss does not improve for $3$ consecutive epochs, with a minimum learning rate threshold of $1.10^{-6}$. This method aids in preventing overfitting and ensures the model converges to an optimal solution.
We employ Binary Cross-Entropy (BCE) as the loss function for our classification model, defined as:

\begin{equation}
\text{BCE} = -\frac{1}{n} \sum_{i=1}^{n} \left[y_i \log\left(\hat{y}_i\right) + \left(1 - y_i\right) \log\left(1 - \hat{y}_i\right)\right],
\end{equation}

where \(y_i\) is the ground-truth label (\(0\) for sound, \(1\) for lame) and \(\hat{y}_i\) is the model’s predicted probability for the positive class. The model was trained on a maximum of 100 epochs with a batch size of 64. The dataset was divided into three distinct sets to ensure balanced evaluation: the training set comprising 60\% of the total data, the validation set accounting for 20\% of the data, and the test set representing the remaining 20\%. This distribution allows for effective model training, hyperparameter tuning, and unbiased performance evaluation.

\subsection{Model Evaluation}

We extended the methodology across all gaits—trot, walk, and canter—to verify its robustness and confirm that our initial approach was correct. For model evaluation, we tested the model on the test dataset to assess its performance in unseen context. After segmenting the data into individual strides, each stride is classified by the model as either sound or lame. To determine an overall performance metric for each session, we calculate the classification score as the proportion of strides labeled as lame out of the total number of strides analyzed in the chosen gait. This stride-level approach offers a more granular view of the model’s performance, reflecting real-world clinical scenarios where consistent detection of abnormal strides across a session is paramount for accurately diagnosing lameness. Also, a custom decision threshold is employed rather than relying on the conventional default of 0.5. This approach is motivated by the inherent imbalance often observed in gait data, where the cost of misclassifying an abnormal stride (a false negative) can be much higher than that of a false positive. To determine the optimal threshold, we compute the precision–recall curve using the model’s probability outputs on the validation set. For each candidate threshold, we calculate the corresponding precision and recall values and then derive the F1 score. The threshold that maximizes the F1 score is selected as the optimal cutoff. This strategy ensures that the model achieves a balanced trade-off between precision and recall, thereby enhancing its overall performance in detecting subtle abnormalities in horse gait.

\section{Results}

We found that the trot gait was the most effective for lameness detection, achieving a classification accuracy of 75.5\%, surpassing the performance observed for the canter (\(74\%\)) and walk (\(53\%\)) gaits, as shown in Fig.~5. While canter yielded promising results, the approach was set aside due to insufficient data from abnormal sessions. Specifically, only six such sessions were available, leading to a recall of 0.66. Fig.~6 illustrates that the trot model accurately classified 6{,}470 out of 8{,}237 abnormal strides, demonstrating a strong sensitivity to detect lameness. These results underscore the model’s reliability and its potential for practical applications in equine health monitoring.

\begin{figure}[h!]
\centering
\begin{tikzpicture}
\begin{axis}[
    width=8cm,
    height=7cm,
    ymin=0,
    ymax=100,
    ybar,
    bar width=2em,
    axis on top,
    axis x line=bottom,
    axis y line=left,
    symbolic x coords={Walk,Trot,Canter},
    xtick=data,
    enlarge x limits=0.3,
    ylabel={\textbf{Test Accuracy (\%)}},
    xlabel={\textbf{Gait}},
    ticklabel style={font=\small},
    label style={font=\small\bfseries},
    title style={yshift=-1.5ex},
    grid=none % removes vertical/horizontal lines
]
\addplot [
    draw=none, 
    fill=blue!70
] coordinates {
    (Walk,53) (Trot,75) (Canter,74)
};

\end{axis}
\end{tikzpicture}
\caption{Comparison of test accuracy on strides for different gaits}
\label{fig:gait-accuracy}
\end{figure}
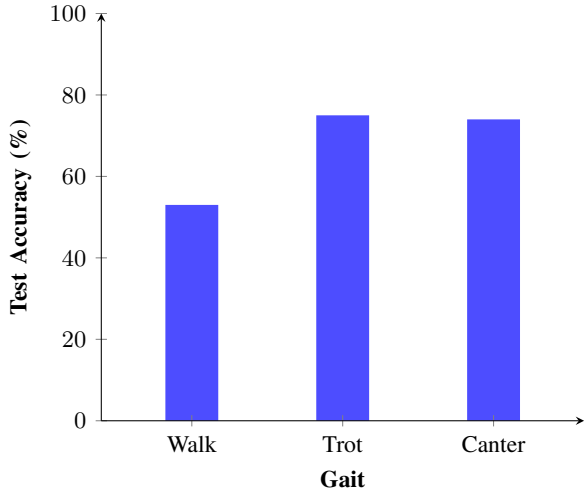

When examining the confusion matrix, our objective is to maximize the values along the diagonal, which represent correct classifications of both sound and lame strides. Conversely, the anti-diagonal corresponds to misclassifications—false positives and false negatives. Of these, false negatives are particularly critical, as they indicate a horse may be lame yet goes undetected. Minimizing such cases is essential for ensuring prompt intervention and maintaining the horse’s health and performance.

Because our main focus is session-level classification, we include an additional confusion matrix at this scale. Fig.~7 shows this matrix, illustrating the overall performance of our approach at the session level. Specifically, out of 32 total sessions, the model correctly classified 29, with no false negatives—including 13 abnormal sessions identified correctly. This 90\% session-level accuracy underlines the effectiveness of our stride-to-session aggregation strategy, further confirming the practical value of our method in routine equine health monitoring.

\begin{figure}[h]
\centering
\includegraphics[scale=0.4]{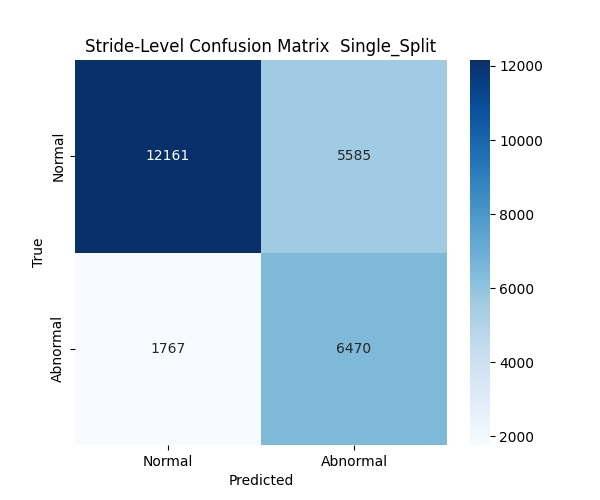}
\caption{Confusion matrix of the 1D CNN model on trot strides}
\end{figure}

\begin{figure}[h]
\centering
\includegraphics[scale=0.4]{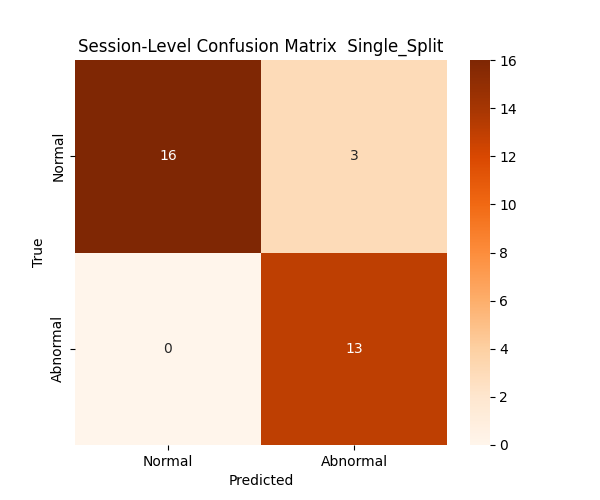}
\caption{Confusion matrix of the 1D CNN model on sessions}
\end{figure}

\section*{Conclusion}
In conclusion, our study introduces a novel approach to detecting early-stage lameness in horses through a supervised learning framework that utilizes data from a single wearable sensor. This innovative method significantly simplifies the data collection process and reduces the overall cost and complexity associated with traditional multi-sensor and video-based approaches. By leveraging a one-dimensional convolutional neural network, our system effectively captures gait dynamics from both sound and lame horses, achieving a training accuracy of 90\% across 32 sessions with zero false negatives. This robust performance enables accurate classification of lameness and supports reliable, data-driven decision-making.
Compared to traditional methods, which often rely on subjective visual assessments or require cumbersome equipment such as force plates and high-speed cameras, our approach offers a more practical and scalable solution. The use of a single sensor not only minimizes intrusiveness but also ensures ease of deployment in various environments, making it accessible for routine use in equine health management. The promising results obtained highlight the potential of our method to revolutionize equine health monitoring, ultimately contributing to better outcomes for horses and their caregivers.

\section*{Future Work}
Building on these findings, several avenues remain open for further investigation. First, we plan to explore methods for evaluating the model’s performance on more dynamic activities that impose varied stresses on a horse’s limbs. Second, we aim to perform a temporal analysis for individual horses, examining how lameness indicators evolve over time within the same animal. Finally, we are committed to enhancing stride-level classification while maintaining the high quality of session-level classification.
In addition, our analysis of the temporal distribution of detected anomalies revealed that they are not confined to specific segments of the gait cycle but instead appear to be spread continuously across the strides. This observation suggests that the model may be capturing ongoing, subtle deviations in stride dynamics rather than isolated phases of abnormality. These findings invite further inquiry into the underlying temporal patterns and their potential implications for early and consistent lameness detection, opening new avenues for future research into the dynamics of gait irregularities in horses.

\vspace{9pt}

\onecolumn
\appendix{}

\begin{figure}[h]
\centering
\includegraphics[scale=0.6]{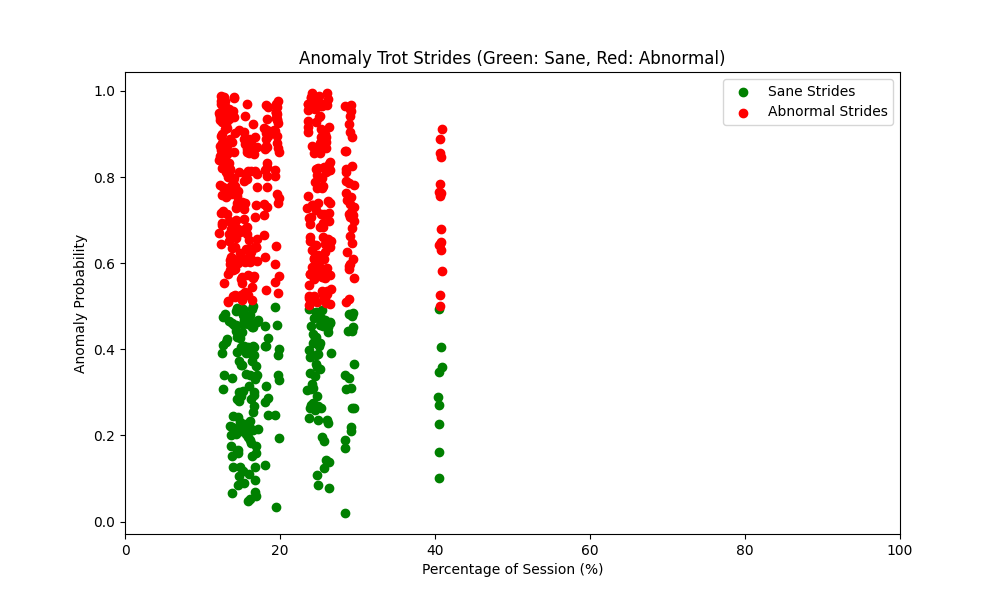}
\end{figure}
\vspace{-1cm}

\begin{figure}[h]
\centering
\includegraphics[scale=0.6]{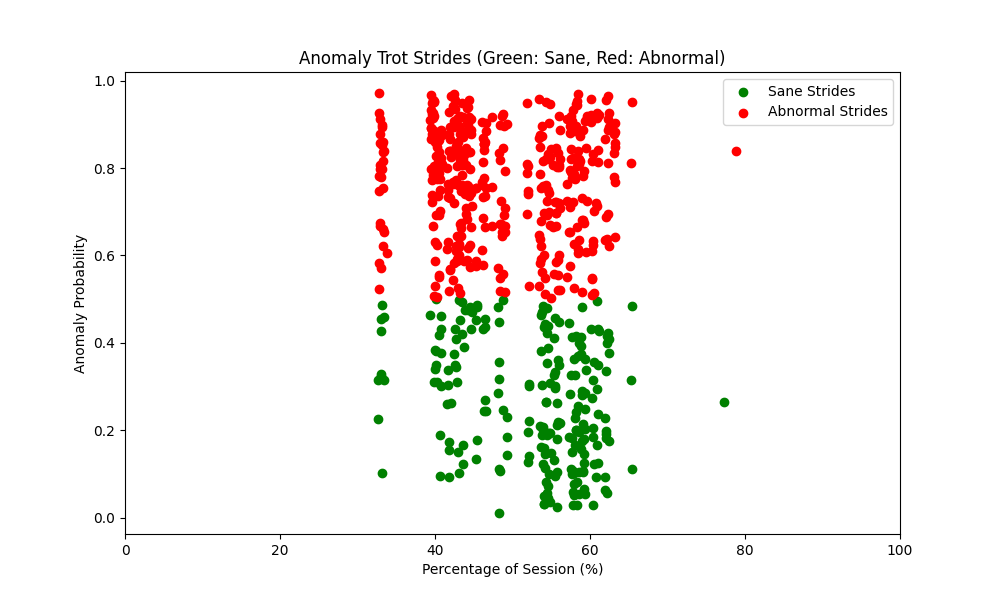}
\caption{Distribution of normal and lame trot strides in two different test sessions classified as lame}
\end{figure}

\end{document}